\documentclass[sn-mathphys]{sn-jnl}

\jyear{2021}%

\theoremstyle{thmstyleone}%
\theoremstyle{thmstyletwo}%

\theoremstyle{thmstylethree}%

\usepackage{amsfonts} 

\begin{document}

\title[Paranasal Anomaly Classification In The Maxillary Sinus]{Multiple Instance Ensembling For Paranasal Anomaly Classification In The Maxillary Sinus}


\author[1,2]{Debayan Bhattacharya}
\author[1]{Finn Behrendt}
\author[2]{Benjamin Tobias Becker}
\author[3]{Dirk Beyersdorff}
\author[4]{Elina Petersen}
\author[5]{Marvin Petersen}
\author[5]{Bastian Cheng}
\author[2]{Dennis Eggert}
\author[2]{Christian Betz}
\author[2]{Anna Sophie Hoffmann*}
\author[1]{Alexander Schlaefer*}

\affil[1]{Institute of Medicial Technology and Intelligent Systems, Technische Universitaet Hamburg, Germany}
\affil[2]{ 
Department of Otorhinolaryngology, Head and Neck Surgery and Oncology}
\affil[3]{ 
Clinic and Polyclinic for Diagnostic and Interventional Radiology and Nuclear Medicine}
\affil[4]{ Population Health Research Department, University Heart and Vascular Center}
\affil[5]{ Clinic and Polyclinic for Neurology}
\affil{University Medical Center Hamburg-Eppendorf, Hamburg, Germany}


\footnotetext{*Equal contribution}
\footnotetext{For correspondence send email to: debayan.bhattacharya@tuhh.de,d.bhattacharya@uke.de}


 \abstract{

 \textbf{Purpose:} Paranasal anomalies are commonly discovered during routine radiological screenings and can present with a wide range of morphological features. This diversity can make it difficult for convolutional neural networks (CNNs) to accurately classify these anomalies, especially when working with limited datasets. Additionally, current approaches to paranasal anomaly classification are constrained to identifying a single anomaly at a time. These challenges necessitate the need for further research and development in this area. \\
\textbf{Methods:} In this study, we investigate the feasibility of using a 3D convolutional neural network (CNN) to classify healthy maxillary sinuses (MS) and MS with polyps or cysts. The task of accurately identifying the relevant MS volume within larger head and neck Magnetic Resonance Imaging (MRI) scans can be difficult, but we develop a straightforward strategy to tackle this challenge. Our end-to-end solution includes the use of a novel sampling technique that not only effectively localizes the relevant MS volume, but also increases the size of the training dataset and improves classification results. Additionally, we employ a multiple instance ensemble prediction method to further boost classification performance. Finally, we identify the optimal size of MS volumes to achieve the highest possible classification performance on our dataset.  \\
 \textbf{Results:} With our multiple instance ensemble prediction strategy and sampling strategy, our 3D CNNs achieve an F1 of 0.85 \(\pm\) 0.09 whereas without it, they achieve an F1 of 0.70 \(\pm\) 0.13.\\
 \textbf{Conclusion:} We demonstrate the feasibility of classifying anomalies in the MS. We propose a data enlarging strategy alongside a  novel ensembling strategy that proves to be beneficial for paranasal anomaly classification in the MS.}

\keywords{Paranasal anomaly, maxillary sinus, CNN, classification}



\maketitle

\section{Introduction}\label{sec1}

The paranasal sinuses are air-filled chambers in the human body that serve as extensions of the nasal cavities and are located within specific bones, such as the frontal, sphenoid, ethmoid, and maxillary bone \cite{martini2012human}. These sinuses are prone to developing pathologies, such as retention cysts \cite{Bal2014-uw} and polyps, which can be identified through routine radiological screenings. In fact, research has shown that the MS are the most commonly and severely affected by these anomalies \cite{Varshney2015-qi}. However, these findings are often incidental, meaning they are unrelated to the patient's primary clinical indications. As a result, paranasal anomalies present several challenges for healthcare professionals in the clinical setting \cite{Hansen.2014}. Multiple studies have been conducted to assess the prevalence of these anomalies in the general population, highlighting the importance of understanding and addressing these pathologies \cite{Tarp.2000,Rak.1991,Stenner.2014,Rege.2012,Cooke.1991}.

Accurate diagnosis of paranasal inflammations is decisive for effective patient care in the healthcare system. Medical professionals often use CT and MRI scans to examine the head and neck area, including the skull base, orbits, and infracranial spaces, in order to assess the local extent of these conditions \cite{Brierley.2017}. The use of 3D information is crucial for correctly identifying paranasal anomalies. Misdiagnosis of these abnormalities can cause unnecessary stress for patients and add unnecessary costs to the healthcare system \cite{Gutmann.2013}. A retrospective study found that inverted papillomas were misdiagnosed as nasal polyps in 8.4\% of cases, and malignant tumors were also misdiagnosed as nasal polyps in 5.63\% of cases \cite{Ma.2012}. To improve diagnosis accuracy and reduce the workload of clinicians, the use of deep learning methods may be beneficial. However, it is important to keep in mind that the paranasal sinuses are highly variable in terms of anatomy \cite{Papadopoulou2021-ch}, requiring careful consideration when using deep learning methods to ensure reliable and accurate diagnoses.

Deep learning has proven to be a valuable tool in the screening of paranasal pathology, with various studies using convolutional neural networks (CNNs) to classify different types of sinusitis \cite{Jeon.2021,Kim.2019} and distinguish between inverted papilloma tumors and  inverted papilloma–associated squamous cell carcinoma \cite{Liu.2022}. Contrastive learning with regular cross-entropy loss has also been used to classify between healthy and anomalous MS \cite{10.1007/978-3-031-16437-8_41}. Additionally, the classification of anomalies in the MS has been approached as an unsupervised anomaly detection task \cite{https://doi.org/10.48550/arxiv.2211.01371}. However, the large anatomical variations of the MS and the morphological variation of the pathologies can make the classification challenging, as deep learning networks may overfit on the training and validation sets. It is also important to note that the high confidence of deep learning models can be misleading \cite{https://doi.org/10.48550/arxiv.1606.06565,DBLP:journals/corr/VarshneyA16}, highlighting the need for caution and careful evaluation in their use in clinical practice. 

Our proposed solution for classifying paranasal anomalies in the MS involves an end-to-end approach that distinguishes between normal and anomalous MS. To achieve this, we first propose a dataset extraction and enlargement strategy that localises the relevant MS area and utilizes an implicit transnational augmentation to obtain additional MS volumes. This is done by sampling three-dimensional coordinates of the approximate centroid of the left and right MS from a Gaussian distribution, and using these coordinates to extract MS volumes. This sampling strategy allows us to increase the size of the dataset while also enabling the extraction of multiple partially overlapping instances of the MS for classification. In addition, we propose a multiple instance ensemble prediction approach that considers various overlapping potential candidate volumes from a single patient's head and neck MRI, and uses a 3D CNN to classify all of these volumes into one of two classes: normal or anomaly. The final prediction for a patient is the average prediction of the multiple MS volumes extracted. Finally, we perform experiments to find the optimal MS volume size that leads to best classification performance and comment on the careful selection of the size of the extracted volume in order to optimize performance. Altogether, by combining our dataset enlargement strategy with our ensemble prediction approach, our 3D CNN has the potential to accurately classify paranasal anomalies that exhibit a wide range of morphological variations and locations.

\section{Methods}

\begin{figure}
    \centering
    \includegraphics[width=1.0\textwidth]{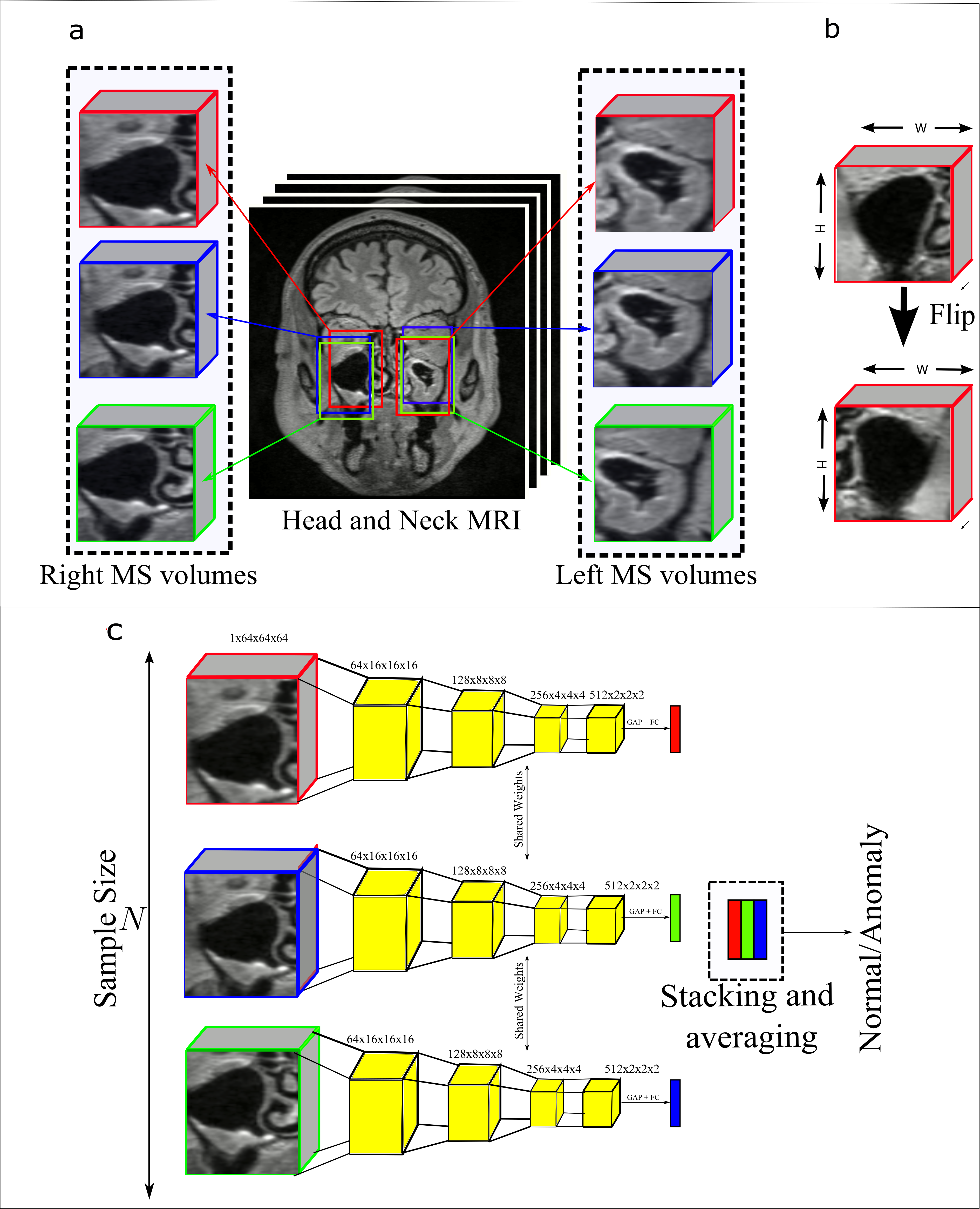}
    \caption{(a) Illustration of our MS volume extraction strategy showing 3 MS volumes for left and right MS each. (b) Flipping of the coronal plane of the right MS (c) Illustration of our multiple instance ensemble prediction strategy used during inference. GAP denotes Global Average Pooling and FC denotes Fully Connected Layer }
    \label{fig:my_label}
\end{figure}
\textit{Dataset}: As part of the Hamburg City Health Study (HCHS) \cite{Jagodzinski2020}, cMRIs of participants (45-74 years) were recorded for neuroradiological assessment. These scans were obtained at the University Medical Center Hamburg-Eppendorf and feature fluid attenuated inversion recovery (FLAIR) sequences in the NIfTI format. The dataset comprises 299 patients, with 174 exhibiting healthy left and right MS and 125 exhibiting at least one MS having a polyp or cyst pathology. The diagnoses were confirmed by two ear, nose, and throat (ENT) surgeons and one ENT specialized radiologist. The anomalies under consideration in this study include polyps and cysts.  MS exhibiting these anomalies are grouped into "anomalous" class and MS without these anomalies are grouped into "normal" class.

\textit{Dataset  Preprocessing and  MS  volume  extraction} : Each MRI in the study has a resolution of 173x319x319 voxels, with a voxel size of 0.53 mm x 0.75 mm x 0.75 mm. To ensure consistency across all of the head and neck MRI scans in our study, we apply a process of rigid registration. This involves selecting one MRI as a fixed volume and registering other MRIs with respect to the fixed volume. Then, we resample the head and neck MRI to a dimension of 128 x 128 x 128 voxels. 

To increase the size of the dataset and be able to use multiple instances of MS volumes for our ensemble prediction, we extracted multiple sub-volumes of left and right MS from individual head and neck MRI scans.  This was done by manually recording the centroid locations of the left and right MS of 20 patients, and using these coordinates to compute the mean and standard deviation of the centroid locations. These values are denoted as \(\mu(x),\mu(y),\mu(z)\) and \(\sigma(x),\sigma(y),\sigma(z)\) for the mean and standard deviation, respectively. We then initialize Gaussian distributions - \(\mathcal{N}(\mu(x),\sigma^{2}(x))\),\(\mathcal{N}(\mu(y),\sigma^{2}(y))\),\(\mathcal{N}(\mu(z),\sigma^{2}(z))\) - and use these distributions to sample centroid locations for MS volumes in the head and neck MRI. It is worth noting that the mean and standard deviation of the left and right MS volumes are different, resulting in a total of six Gaussian distributions in practice. We sample \(N\) left MS volumes and \(N\) right MS from each head and neck MRI where \(N\) is the sample size. For our experiments, \(N \in \{1,5,10,15,20\}\).  An illustration of our sampling method is shown in figure 1 (a).  We extract MS volumes of multiple sizes namely, 25×25×25, 30×30×30, 35×35×35, 40×40×40, 45×45×45.  The extracted MS volumes are finally resampled to a resolution of 64×64×64 for the 3D CNN. To make the right and left MS appear more symmetrical, we horizontally flip the coronal planes of the right MS to give it the appearance of the left MS volume.

\textit{Training, validation and test splits}:
If we sample with \(N=1\), we end up with 327, 37 and 41  MS volumes in the training, validation and test set respectively. The training validation and test split increase by a factor of \(2N\) with respect to the sample size \(N\). 32\% of the MS volumes in the training, validation and test sets are anomalous MS volumes. We perform 3-fold cross validation experiments with all the methods. 

\textit{Implementation Details}
We implement a 3D CNN using ResNet18 \cite{Hara.25.08.2017} with 4 stages of 3D residual blocks (channel dimensions 64, 128, 256, 512). Our models are trained for 100 epochs with a batch size of 16, a learning rate of 0.0001, and Adam optimization. If the validation loss did not improve for 5 epochs, the learning rate is reduced by a factor of 10. We use PyTorch and PyTorch Lightning to build our models.

\textit{Deep Learning method}: To classify the MS volume into normal or anomaly class, we use a 3DResNet \cite{Hara.25.08.2017}\footnote{https://github.com/kenshohara/3D-ResNets-PyTorch/blob/master/models/resnet.py}. Let us denote the classifier as \(f(.)\) and the MRIs as \(X \in R^{H \times W \times D} \). From each MRI, we extract \(N\) left MS volumes and \(N\) right MS volumes. Altogether, we extract 2\(N\) MS volumes from \(X \in R^{H \times W \times D} \). Let us denote the MS volumes as \(x \in R^{P \times P \times P} \). Here, \(P\) denotes the size of the MS volume such that \(P \in \{25,30,35,40,45\}\). Further, our labels \(y \in \{0,1\}\) represent normal and anomaly class. The anomaly class is the positive class for our use-case. As a baseline, we consider 3DResNet models that do not use our multiple instance ensemble strategy for inferring on the test set. 

\textit{Multiple Instance Ensemble Prediction Strategy}: 
Let us denote the extracted MS volumes from a single MRI \(x_{i} \in R^{P \times P \times P}\) where \(i\) denotes the \(i-th\) MS volume extracted from either the left or right MS area of the MRI.  When making a prediction, we average the softmax scores of classifier \(f(.)\) from the multiple MS volumes \(x_{i}\). Formally, 
\begin{equation*}
    \hat{y} =  \frac{1}{N}\sum_{i=1}^{N} softmax(f(x_{i})) 
\end{equation*}

\section{Results}\label{sec2}

We plot the mean and standard deviation of the Area Under Precision Recall Curve (AUPRC) and F1 score. Both these metrics, are useful especially in imbalanced scenarios which is our case. From the table \ref{tab1}, we observe that with the increase in the sample size \(N\), we get a consistent increase in all the reported metrics until \(N=15\) after which we get a decrease in all the metrics. Further, for all the cases, we see that using multiple instance ensemble strategy is beneficial for MS anomaly classification and leads to boost in classification metrics. 

\begin{figure}
    \centering 
    \includegraphics[width=0.8\textwidth]{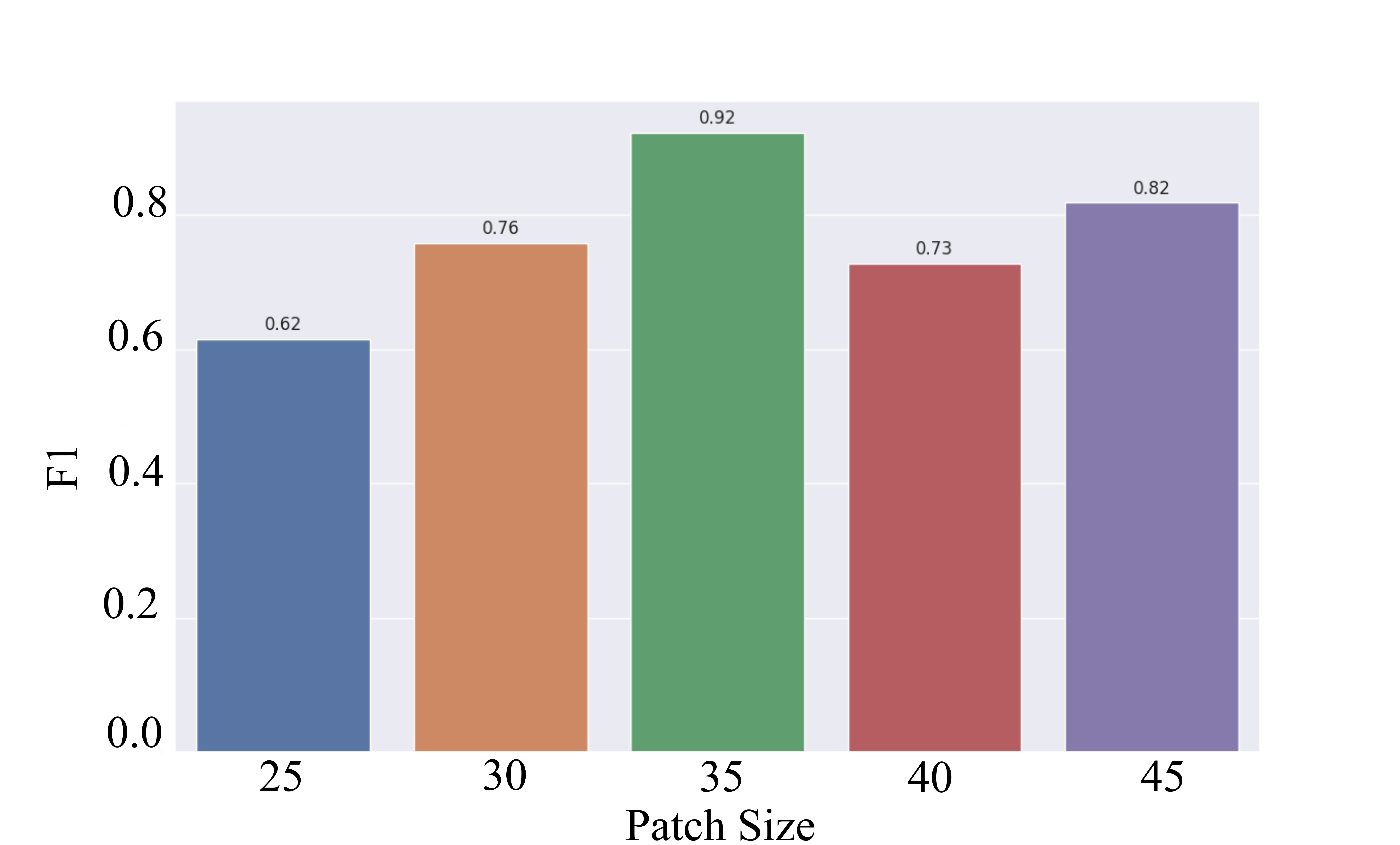}
    \caption{F1 scores vs Patch Size \(P\)}
    \label{fig:2}
\end{figure}

Further, looking at figure \ref{fig:2}, we can see the influence of MS volume size to the parnasal classification task. Note, we set \(N=15\) for this experiment and use our multiple instance ensemble prediction strategy. This highlights that patch size plays an important role in boosting the paranasal anomaly classification performance.  Our experiments indicate that that the optimal patch size
for our dataset is \(P = 35\).

\begin{table}[h]

\begin{center}
\begin{minipage}{174pt}
\caption{Result of our experiments}\label{tab1}%
\begin{tabular}{@{}llll@{}}

\toprule
N & Ensemble Prediction  & AUPRC & F1 \\
\midrule
1    &   &0.80$\pm$0.12 & 0.70$\pm$0.13 \\
\hline
5    &    & 0.85$\pm$0.03 & 0.77$\pm$0.10  \\
5    & \checkmark   &  0.87$\pm$0.04 & 0.76$\pm$0.10  \\
\hline
10    &   & 0.85$\pm$0.04  & 0.75$\pm$0.08   \\
10    & \checkmark   & 0.89$\pm$0.05  & 0.79$\pm$0.10  \\
\hline
15    &    & 0.88$\pm$0.07  & 0.81$\pm$0.12  \\
15    & \checkmark   & \textbf{0.92$\pm$0.06}  & \textbf{0.85$\pm$0.09}  \\
\hline
20    &    & 0.87$\pm$0.04  & 0.77$\pm$0.05  \\
20    & \checkmark   & 0.91$\pm$0.02  & 0.78$\pm$0.07  \\
\botrule
\end{tabular}
\footnotetext{Patch size \(P\) = 35 for all the experiments}
\end{minipage}
\end{center}
\end{table}

\section{Discussion}\label{sec12}

\begin{figure}
    \centering
    \includegraphics[width=0.8\textwidth]{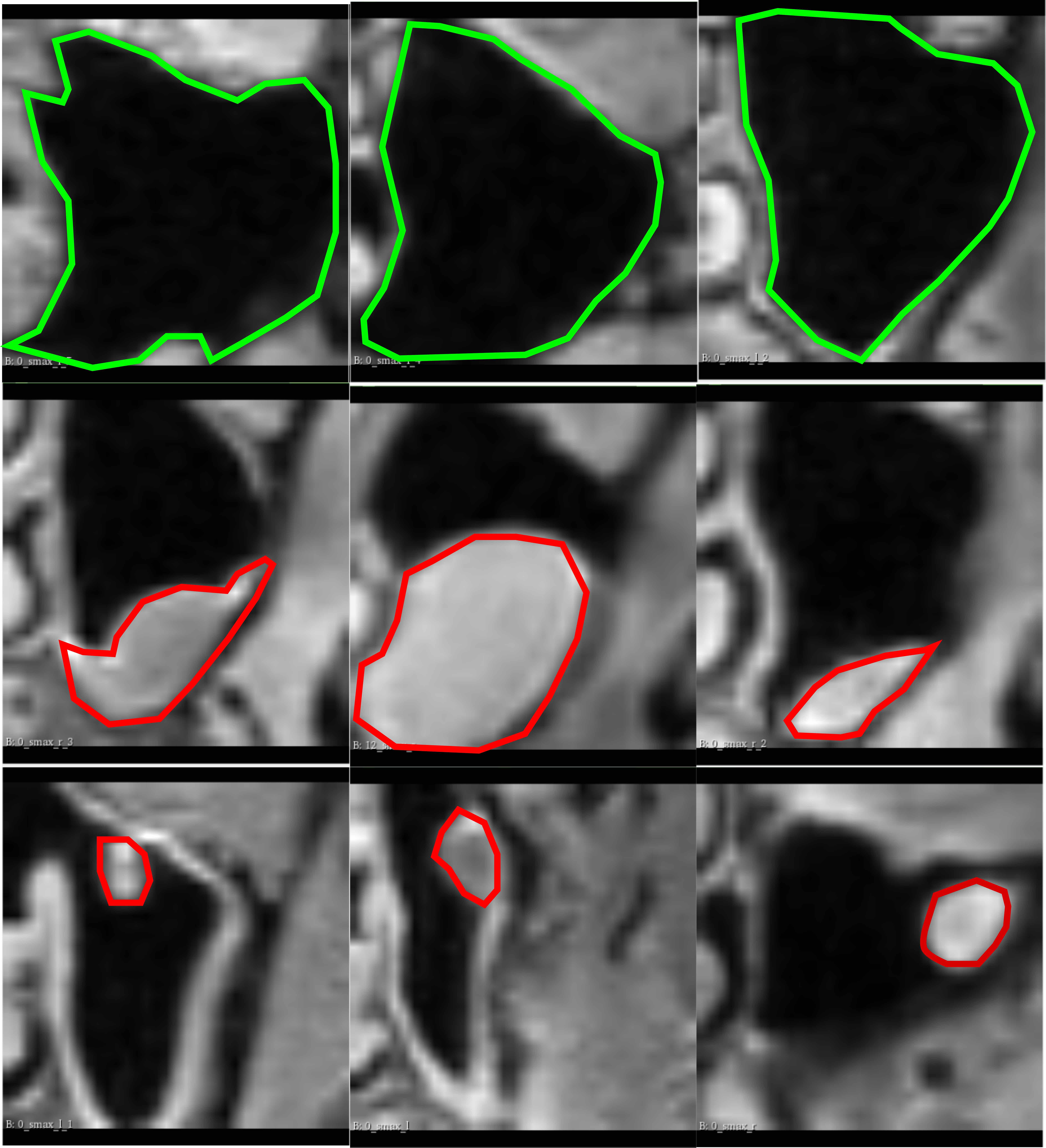}
    \caption{The coronal planes of the sampled MS volumes. The green contours in the first row represent normal MS anatomy. The red contours enclose masses that represent cysts and polyps in the second and third row respectively, demonstrating the variety of appearances and morphological variations of these anomalies within the MS.}
    \label{fig:my_label}
\end{figure}
In order to accurately classify anomalies in the MS, it is necessary to first extract the relevant volumes from a head and neck MRI. While deep learning methods using 3D object detection have been suggested for this purpose \cite{Kern2020}, they require the manual labeling of the MS location by specialized clinicians as ground truth data, which may not always be feasible. As an alternative, we propose a method that extracts MS volumes by modeling the centroid location of the MS volumes using a gaussian distribution, which is both efficient and does not require significant human effort. However, it is possible that our method may not extract a sub-volume that fully encompasses the MS if the patch size is too small. To address this issue, we extract multiple sub-volumes. Our analysis shows that as the sample size increases, the classification metrics improve, although there is a decrease in the F1 score for a sample size of 20 compared to 15. This may be due to the inclusion of redundant MS volumes leading to overfitting and a loss in generalizability. It is therefore important to carefully select the appropriate sample size for this task. Additionally, using an ensemble strategy that averages the scores from multiple instances of the MS leads to a further improvement in classification metrics. The improvement in our classification metrics can be attributed to the incorporation of implicit test-time augmentation during inference on the test set. By sampling multiple overlapping MS volumes, we have MS volumes which have transnational offsets with respect to one another, resulting in better performance. These findings demonstrate the utility of our proposed method for the classification of paranasal anomalies in the MS. 

\begin{figure}
    \centering
    \includegraphics[width=0.8\textwidth]{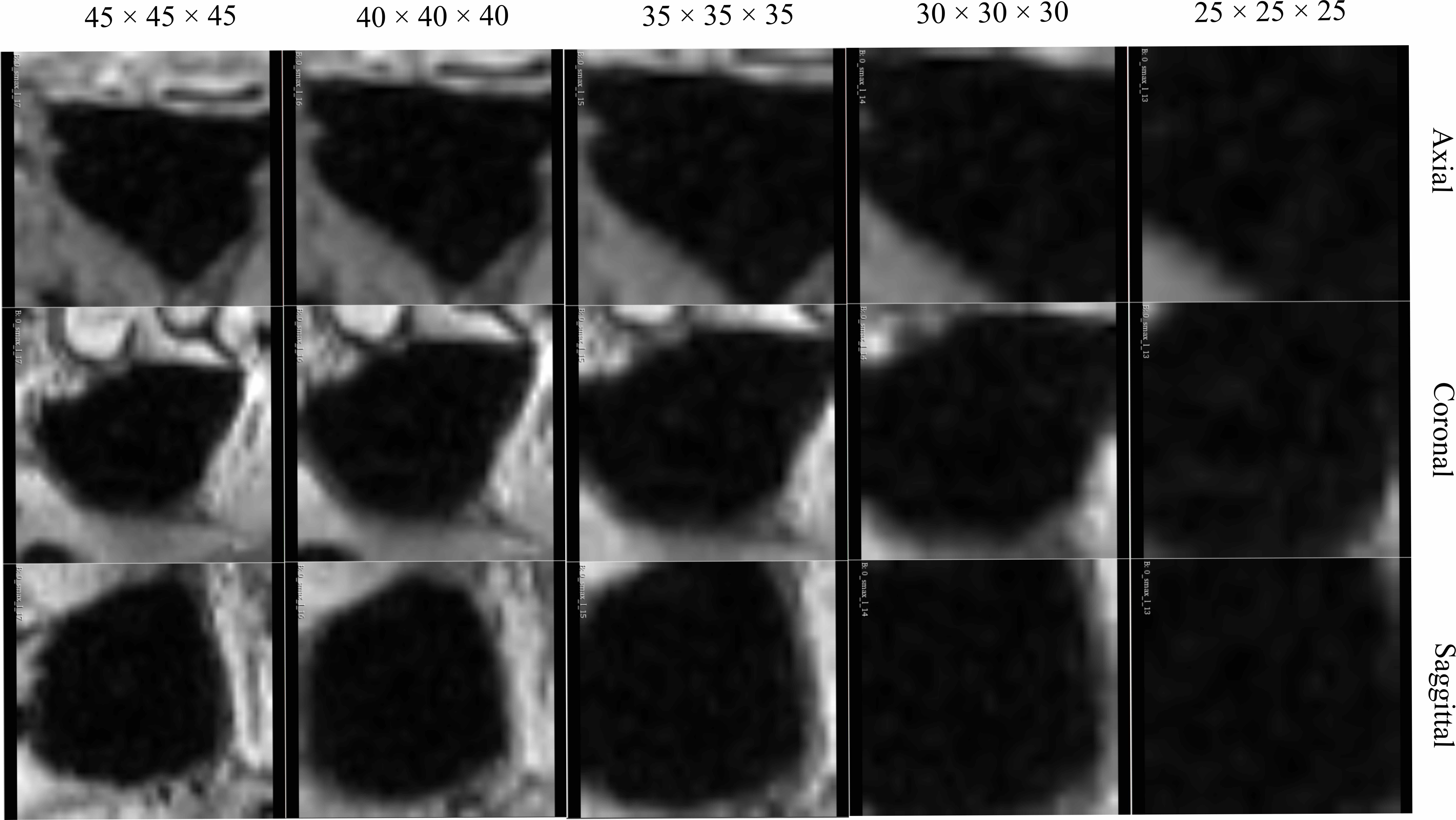}
    \caption{ Slices from the axial, coronal and saggital slices of extracted healthy MS volume with different patch sizes. }
    \label{fig:maxillarysinuses}
\end{figure}

Finally, the size of the extracted MS volume is a crucial factor in the success of paranasal anomaly classification. A volume that is too small may fail to fully capture the pathology or may only partially extract the MS from the head and neck MRI. On the other hand, a volume that is too large may include surrounding anatomies that add unnecessary and irrelevant information, hindering the classification task. This can be seen in figure \ref{fig:maxillarysinuses}. To address this trade-off, we evaluated the classification performance of our 3D CNN on MS volumes of various sizes, including 25 x 25 x 25, 30 x 30 x 30, 35 x 35 x 35, 40 x 40 x 40, and 45 x 45 x 45 voxels. Our results showed that as the volume size increased, the F1 score also increased, but only until a volume size of 35 x 35 x 35. After this point, the F1 score decreased. This suggests that small volumes are unable to fully capture the MS and may miss important anomalies, while larger volumes negatively impact classification performance due to the inclusion of unnecessary surrounding structures. These findings highlight the importance of carefully selecting the size of the extracted sub-volume to achieve optimal performance.

\section{Conclusion}\label{sec13}

We present a deep learning approach for classifying paranasal anomalies in the maxillary sinus. Our method involves using a multiple instance ensemble prediction strategy to boost performance. To increase the size of the training dataset, we develop a sampling strategy that localises the region of interest and generates multiple instances of MS volumes. We also determine the optimal sample size and investigate the trade-off between patch size and classification performance. While our approach shows promising results, further improvements in the F1 score of our 3D CNN are needed to make it suitable for real-world clinical use. Nevertheless, our work provides a potential solution for paranasal anomaly classification in the maxillary sinus using deep learning.

\backmatter

\bmhead{Ethical approval declarations} The local ethics committee for the State of Hamburg Chamber of Medical Practitioners (Landesärztekammer Hamburg, PV5131) was consulted during the planning of the study and gave their approval for the protocol. The study was also approved by the Data Protection Commissioner for the University Medical Center of the University Hamburg-Eppendorf and the Data Protection Commissioner for the Free and Hanseatic City of Hamburg. It has been registered on ClinicalTrial.gov with the identifier NCT03934957. The procedures and practices followed in the study, including the conduct, evaluation, and documentation, follow Good Clinical Practice, Good Epidemiological Practice, and the ethical principles outlined in the Declaration of Helsinki.
\bmhead{Conflicts of Interest}

Debayan Bhattacharya states no conflict of interest. Finn Behrendt states no conflict of interest. Dirk Beyersdorff states no conflict of interest. Elina Petersen states no conflict of interest. Marvin Petersen states no conflict of interest. Bastian Cheng states no conflict of interest. Dennis Eggert states no conflict of interest. Christian Betz states no conflict of interest. Anna Sophie Hoffmann states no conflict of interest. Alexander Schlaefer states no conflict of interest.

\bmhead{Acknowledgments}

This work has not been submitted for publication anywhere else. This work is funded partially by the i3 initiative of the Hamburg University of Technology. The authors also acknowledge the partial funding by the Free and Hanseatic City of Hamburg (Interdisciplinary Graduate School) from University Medical Center Hamburg-Eppendorf. This work was partially funded by Grant Number KK5208101KS0 (Zentrales Innovationsprogramm Mittelstand, Arbeitsgemeinschaft industrieller Forschungsvereinigungen).

\bibliography{sn-bibliography}


\end{document}